\documentclass[journal]{IEEEtran}

\usepackage{cite}
\usepackage{algorithm}
\usepackage{graphicx}
\usepackage{amsmath}
\usepackage{algpseudocode}
\usepackage{subcaption}
\usepackage{array}
\usepackage{booktabs}
\usepackage{tabularx}
\usepackage{url}

\makeatletter
\renewcommand{\paragraph}[1]{%
  \textbf{#1:}
}
\makeatother

\newcommand{\norm}[1]{\left\lVert#1\right\rVert}
\newcommand{\Design}[0]{\textsc{ACORN-IDS}}

\begin{document}

\title{ACORN-IDS: Adaptive Continual Novelty Detection for Intrusion Detection Systems}

\author{Sean~Fuhrman,
        Onat~Gungor, and~Tajana~Rosing
\thanks{Department of Computer Science and Engineering, University of California, San Diego, CA 92093 USA (email: \{stfuhrma, ogungor, tajana\}@ucsd.edu).}}


\maketitle

\begin{abstract}
Intrusion Detection Systems (IDS) must maintain reliable detection performance under rapidly evolving benign traffic patterns and the continual emergence of cyberattacks, including zero-day threats with no labeled data available. However, most machine learning–based IDS approaches either assume static data distributions or rely on labeled attack samples, substantially limiting their applicability in real-world deployments. This setting naturally motivates continual novelty detection, which enables IDS models to incrementally adapt to non-stationary data streams without labeled attack data. In this work, we introduce \Design{}, an adaptive continual novelty detection framework that learns exclusively from normal data while exploiting the inherent structure of an evolving unlabeled data stream. \Design{} integrates a continual feature extractor, trained using reconstruction and metric learning objectives with clustering-based pseudo-labels, alongside a PCA-based reconstruction module for anomaly scoring. This design allows \Design{} to continuously adapt to distributional shifts in both benign and malicious traffic. We conduct an extensive evaluation of \Design{} on five realistic intrusion datasets under two continual learning scenarios: (i) Evolving Attacks and (ii) Evolving Normal and Attack Distributions. \Design{} achieves, on average, a 62\% improvement in F1-score and a 58\% improvement in zero-day attack detection over the state-of-the-art unsupervised continual learning baseline. It also outperforms existing state-of-the-art novelty detection approaches while exhibiting near-zero forgetting and imposing minimal inference overhead. These results demonstrate that \Design{} offers a practical, label-efficient solution for building adaptive and robust IDS in dynamic, real-world environments. We plan to release the code upon acceptance. 
\end{abstract}

\begin{IEEEkeywords}
Intrusion Detection Systems, Cybersecurity, Machine learning, Continual learning, Novelty Detection
\end{IEEEkeywords}

%
\IEEEpeerreviewmaketitle

\section{Introduction}

\begin{figure}[t]
    \centering
    \begin{subfigure}[t]{\columnwidth}
        \centering
        \includegraphics[width=\linewidth]{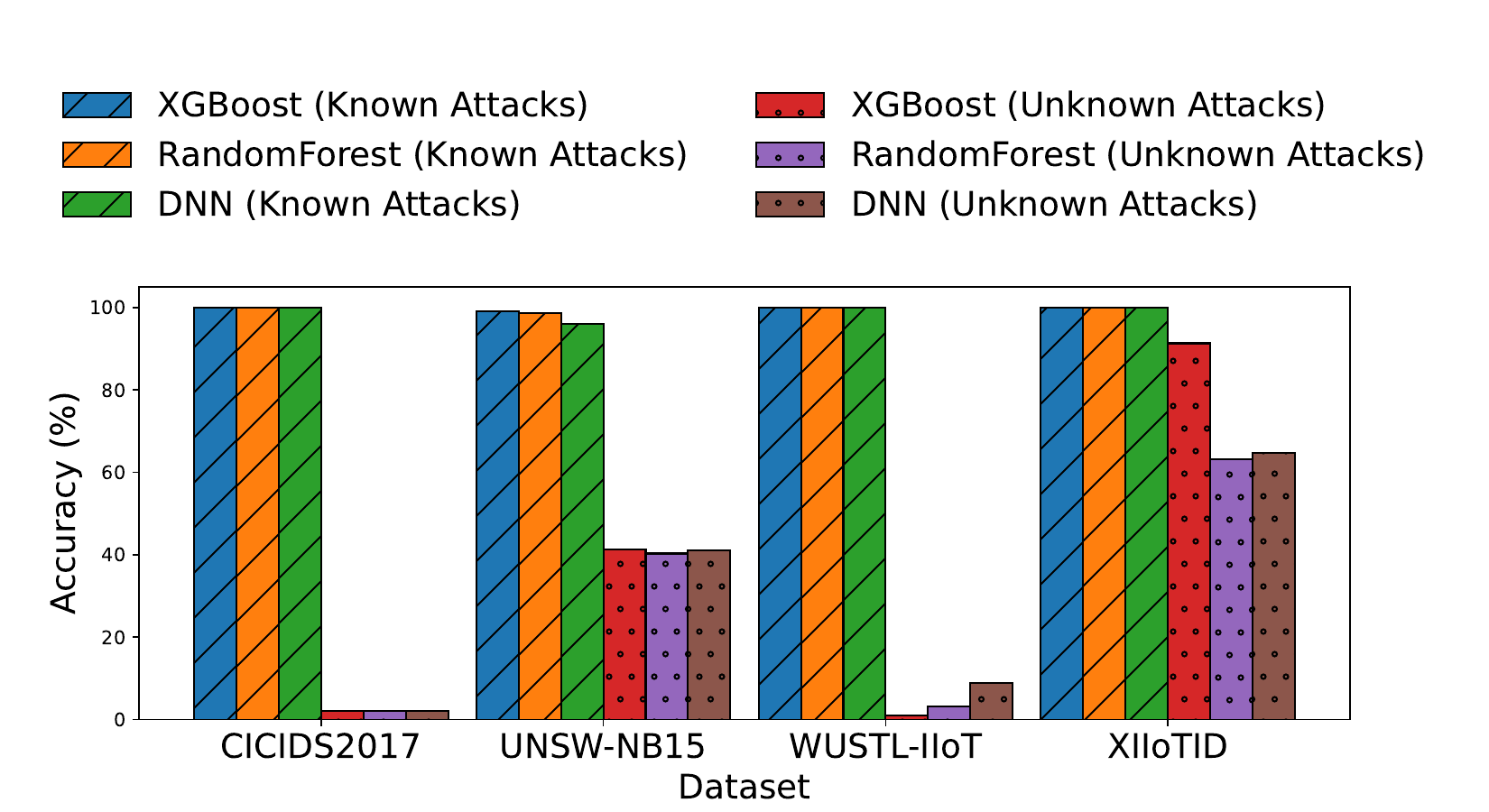}
        \caption{Supervised machine learning intrusion detection performance on known (lined bars) and unknown (dotted bars) attacks.}
        \label{fig:unknown_attacks}
    \end{subfigure}
    \hfill
    \begin{subfigure}[t]{\columnwidth}
        \centering
        \includegraphics[width=\linewidth]{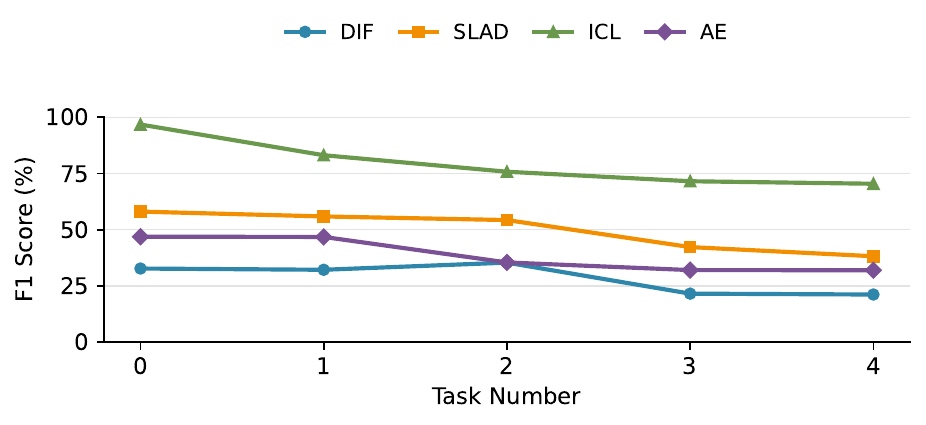}
        \caption{Performance of state-of-the-art novelty detection methods under changing normal data on CICIDS2017~\cite{Sharafaldin2018TowardGA}.} 
        \label{fig:changing_normal}
    \end{subfigure}
    \caption{Motivating example illustrating the effects of distribution shift: (a) evolving attack distributions and (b) evolving normal data distributions. In the absence of continual learning mechanisms, state-of-the-art methods exhibit significant performance degradation under such distributional changes.
    }
    \label{fig:binary_model}
\end{figure}

\IEEEPARstart{C}{ybersecurity} is a cornerstone of modern digital society, essential for safeguarding sensitive data and maintaining trust in connected systems. Given that a new cyberattack occurs approximately every 37 seconds~\cite{aag_cybercrime_stats}, the need for robust and resilient defensive mechanisms has become increasingly urgent. Intrusion Detection Systems (IDS) play a critical role in this effort by monitoring network traffic and user behavior to identify malicious activity~\cite{liao2013intrusion}. While traditional signature-based IDS can detect known threats with high accuracy, they fundamentally fail against zero-day attacks—novel intrusions for which no prior signatures exist~\cite{garcia2009anomaly}.

To address these limitations, machine learning (ML) has been widely adopted for intrusion detection due to its strong detection performance and reduced reliance on expert-crafted rules~\cite{gungor2024roldef}. However, state-of-the-art ML-based IDS methods are typically trained offline and deployed as static models~\cite{verwimp2023continual}. Consequently, their performance degrades over time as both benign traffic patterns and attack strategies evolve~\cite{prasath2022analysis}. Moreover, most ML-based IDS approaches assume access to labeled attack data during training~\cite{kozik2019balanced,karn2021learning,prasath2022analysis,oikonomou2023multi,channappayya2024augmented}, which is rarely feasible in practice. Attack labels are expensive to obtain, highly imbalanced, and fundamentally unavailable for zero-day threats, causing models to overfit to frequent attack patterns while failing to generalize to rare or unseen intrusions~\cite{macas2022survey}. Figure~\ref{fig:unknown_attacks} illustrates this limitation by showing a substantial performance degradation when ML-based IDS models are evaluated on previously unseen attacks.

These challenges motivate the adoption of novelty detection (ND) methods, which learn a model of benign behavior and subsequently identify deviations from this model as potential intrusions~\cite{zong2018deep,verkerken2022towards}. By training exclusively on normal data, ND techniques enable the detection of anomalous and zero-day attacks without requiring labeled attack samples, making them particularly well suited to adversarial settings in which attack distributions evolve rapidly and labeled data are scarce or unavailable~\cite{emerging2023kumar}. Despite these advantages, ND algorithms often exhibit degraded performance in dynamic environments due to issues such as \textit{catastrophic forgetting}, wherein models lose previously learned patterns when adapting to newly observed data~\cite{Faber_2024}. This limitation is illustrated in Fig.~\ref{fig:changing_normal}, which evaluates representative state-of-the-art ND methods—SLAD~\cite{xu2023fascinating}, ICL~\cite{shenkar2022anomaly}, DIF~\cite{xu2023deep}, and an autoencoder-based approach (AE)~\cite{Torabi2023AutoencoderIDS}—under a changing normal data distribution. All methods are trained only once at task~0, after which the normal data distribution shifts across subsequent tasks. As the distribution evolves, performance consistently degrades, highlighting the inability of static ND models to maintain reliable detection in non-stationary environments. 

Continual learning has been proposed as a framework for models that must adapt to an infinite data stream while retaining prior knowledge \cite{de2021continual,wang2024comprehensive}. In this setting, each distribution shift—whether a new type of benign traffic or a new class of attack—constitutes a new task. CL approaches aim to maintain performance across multiple tasks without \textit{catastrophic forgetting}, a capability particularly vital in cybersecurity where both “normal” and “malicious” behavior are constantly changing \cite{li2025citadel}. To illustrate, an IDS trained on attacks such as \textit{worm}, \textit{shellcode}, and \textit{exploits} should also adapt to detect emerging threats like \textit{backdoor} or \textit{fuzzers} \cite{karn2021learning}. An effective IDS must therefore not only adapt its internal representations over time, as in continual learning, but also detect previously unseen behaviors as attacks, even in the absence of labeled attack data. This motivates the study of \textit{Continual Novelty Detection} (CND): a learning paradigm that (i) learns only from normal data, and (ii) adapts continually to evolving distributions of both normal and attack traffic \cite{Faber_2024}. These models can be retrained on new benign data to track the shifting definition of “normal,” while still detecting emerging malicious behavior \cite{FABER_VLAD}. CND unites the strengths of novelty detection and continual learning, providing a principled pathway toward IDS that are robust against both zero-day attacks and distributional drift.

Building on our previous work CND-IDS \cite{fuhrman2025}, we introduce \Design{}, a continual novelty detection framework for intrusion detection (Fig.~\ref{figure:model_diagram}) designed to operate under realistic conditions where attack labels are unavailable and data distributions evolve over time. \Design{} integrates a small amount of labeled normal data with a continuously evolving \emph{unlabeled} data stream, enabling the system to adapt to new and unforeseen patterns without supervision. Unlike traditional novelty detection methods that rely on static normal data \cite{xu2023fascinating, shenkar2022anomaly, xu2023deep} or continual learning approaches that require labeled samples from all classes \cite{kozik2019balanced, oikonomou2023multi, channappayya2024augmented}, \Design{} systematically exploits the structure present in the unlabeled stream to improve both detection and adaptation as the environment changes. \Design{} combines a continual feature extractor (CFE) with a PCA-based reconstruction module for anomaly scoring. The CFE assigns pseudo-labels via K-Means clustering and is trained using a triplet-margin metric-learning objective, allowing it to learn feature embeddings that remain discriminative across tasks. PCA is fit exclusively on labeled clean normal data, while the unlabeled stream contributes to continual representation learning through pseudo-labeling and metric learning. This design enables \Design{} to detect zero-day attacks via reconstruction error thresholding without requiring any labeled attack data, while continually adapting to evolving traffic patterns. The main contributions of this work are summarized as follows:
\begin{itemize}
    \item We propose \Design{}, an adaptive continual novelty detection framework for intrusion detection that operates without labeled attack data and remains effective under evolving benign traffic patterns and emerging cyberattacks. To the best of our knowledge, this is the first continual novelty detection approach that jointly addresses changes in both normal and attack data distributions.
    \item We introduce a continual feature learning mechanism that leverages reconstruction and metric learning objectives, combined with clustering-based pseudo-labels, to enable robust representation learning from an evolving unlabeled data stream.
    \item We integrate a PCA-based reconstruction module for anomaly scoring, allowing \Design{} to detect zero-day attacks while continually adapting to distribution shifts in both benign and malicious traffic.
    \item We conduct extensive evaluations on five realistic intrusion datasets under two continual learning scenarios—Evolving Attacks (EA) and Evolving Normal and Attack Distributions (ENA)—demonstrating that \Design{} achieves, on average, 62\% improvement in F1 scores and 58\% improvement in zero-day attack performance compared to the best performing state-of-the-art unsupervised continual learning baseline, ADCN~\cite{ashfahani2023unsupervised}.
    \item We show that \Design{} outperforms existing novelty detection methods in both thresholded and threshold-free evaluations, while maintaining near-zero forgetting and minimal inference overhead.
\end{itemize}

\section{Related Work}

\textbf{Intrusion Detection Systems.} Intrusion Detection Systems (IDS) are generally classified into two categories: signature-based and anomaly-based. Signature-based IDS identify attacks by matching network traffic against known attack patterns, while anomaly-based IDS employ statistical or machine learning (ML) techniques to model normal behavior and flag deviations as suspicious \cite{gupta2023surveyIDS}. ML-based IDS (ML-IDS), a popular form of anomaly-based IDS, have gained traction due to their strong detection accuracy and reduced reliance on expert-crafted rules \cite{gungor2024rigorous}. The literature spans both supervised methods (e.g., support vector machines, random forests, and deep neural networks \cite{liu2019machine}) and unsupervised approaches (e.g., isolation forest, autoencoders, and principal component analysis \cite{verkerken2022towards}). Despite promising results, most ML-IDS models assume static, independent and identically distributed (i.i.d.) data distributions, an assumption that rarely holds in real-world deployments. In practice, network traffic evolves over time due to changes in user behavior, software updates, and network configurations, while attack strategies continuously adapt in response to deployed defenses, leading to non-stationary and temporally correlated data streams \cite{prasath2022analysis}.

\textbf{Continual Learning for Intrusion Detection.} In practice, benign traffic patterns and attack strategies evolve, causing ML-IDS performance to degrade over time \cite{verwimp2023continual}. Continual Learning (CL) addresses this by enabling models to acquire new knowledge from a non-stationary data stream while retaining previously learned information \cite{wang2024comprehensive}. Several works have adapted CL for IDS. Kozik et al. \cite{kozik2019balanced} extended the Efficient Lifelong Learning Algorithm (ELLA) \cite{ruvolo2013ella} to address class imbalance in cyber attack detection. Karn et al. \cite{karn2021learning} proposed a Learning Without Forgetting (LwF) framework for threat detection, while Kumar et al. \cite{amalapuram2022continual} explored Elastic Weight Consolidation and Gradient Episodic Memory. Other work has explored multi-class CL-IDS frameworks \cite{oikonomou2023multi} and memory replay strategies for balancing classes and improving sample efficiency \cite{channappayya2024augmented}. 
Recent work in IDS-specific CL frameworks further highlight drift-aware adaptation. Zhang et al.~\cite{zhang2024ssf} proposed SSF, which refreshes a fixed memory buffer through strategic sample selection and forgetting to better track concept drift. Zhang et al.~\cite{zhang2024aocids} introduced AOC-IDS, an autonomous online IDS that reduces labeling requirements by leveraging an autoencoder with a contrastive objective. Despite these advances, these studies rely on labeled attack data, which is costly to obtain and unavailable for zero-day threats.

\textbf{Novelty Detection.} Novelty detection (ND) offers an alternative by learning exclusively from normal data, allowing IDS to detect intrusions without requiring labeled attacks \cite{verkerken2022towards}. Methods such as Principal Component Analysis (PCA) and Auto-Encoders (AE)~\cite{li2025safe} detect anomalies through reconstruction, where the reconstruction error of an anomalous sample would be higher. On the other hand, density-based models such as One-Class SVM (OCSVM) \cite{scikit-learn}, Isolation Forest (IF) \cite{scikit-learn}, and Local Outlier Factor (LOF) \cite{scikit-learn} identify low-density or sparsely isolated regions of feature space as potential anomalies. Recently, deep learning approaches have been proposed that utilize different methods to detect anomalies. Deep Isolation Forest (DIF) \cite{xu2023deep} enhances tree-based isolation with deep feature extraction, and Internal Contrastive Learning (ICL) \cite{shenkar2022anomaly} leverages self-supervised contrasting of internal feature statistics to improve sensitivity to subtle deviations in tabular data. Deep Anomaly Detection with Scale Learning (SLAD) \cite{xu2023fascinating} further introduces a scale-learning mechanism to improve the supervision signal from the normal data. While ND methods are well suited for zero-day detection, they struggle under evolving distributions, often suffering from \textit{catastrophic forgetting} when shifts occur in normal data~\cite{Faber_2024}.

\textbf{Continual Novelty Detection.} Recent research has begun to explore continual novelty detection (CND), which integrates ND with CL to handle distribution shifts after deployment. Kim et al.~\cite{kim2023openworld} argued that ND is critical for realistic class-incremental learning. Aljundi et al.~\cite{aljundi2022continual} introduced CND for detecting out-of-distribution data relative to current model knowledge, while Rios et al.~\cite{rios2022incdfm} proposed incremental feature modeling to incorporate novel data into CL systems. However, Faber et al.~\cite{Faber_2024} showed that even ND algorithms themselves degrade under continual settings, underscoring the challenge of dynamic environments. This motivates the development of IDS frameworks that continually adapt using only normal data. Faber et al. proposed VLAD~\cite{FABER_VLAD}, a deep, task-agnostic novelty detection approach that learns latent representations of normal data using variational autoencoders and detects anomalies based on deviations in the learned latent space, without requiring attack samples. Our previous work CND-IDS~\cite{fuhrman2025}, utilized a continual feature extractor alongside PCA-novelty detection to evolve to new attack patterns. \Design{} extends CND-IDS by explicitly addressing evolving normal data distributions in addition to emerging attacks. In contrast to CND-IDS, \Design{} adopts a replay-based strategy that interleaves representative samples from past tasks during training, providing a simpler and more robust mechanism for mitigating catastrophic forgetting.

\section{\Design{} Framework}
\label{framework}
\begin{figure*}[!t]
    \centering
    \includegraphics[width=\linewidth,height=\textheight,keepaspectratio]{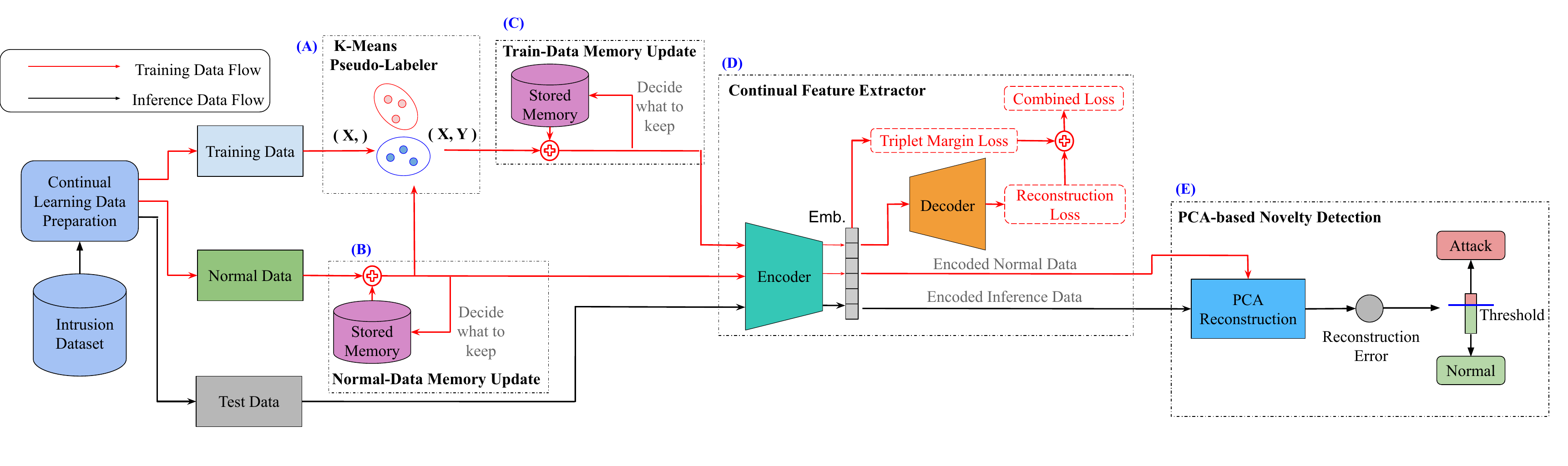}
    \caption{Overview of the \Design{} continual novelty detection framework for intrusion detection systems.}
    \label{figure:model_diagram}
\end{figure*} 

\subsection{Continual Learning (CL) Data Preparation}
\label{CL_data_prep}

\begin{figure}[t]
    \centering
    \includegraphics[width=\columnwidth]{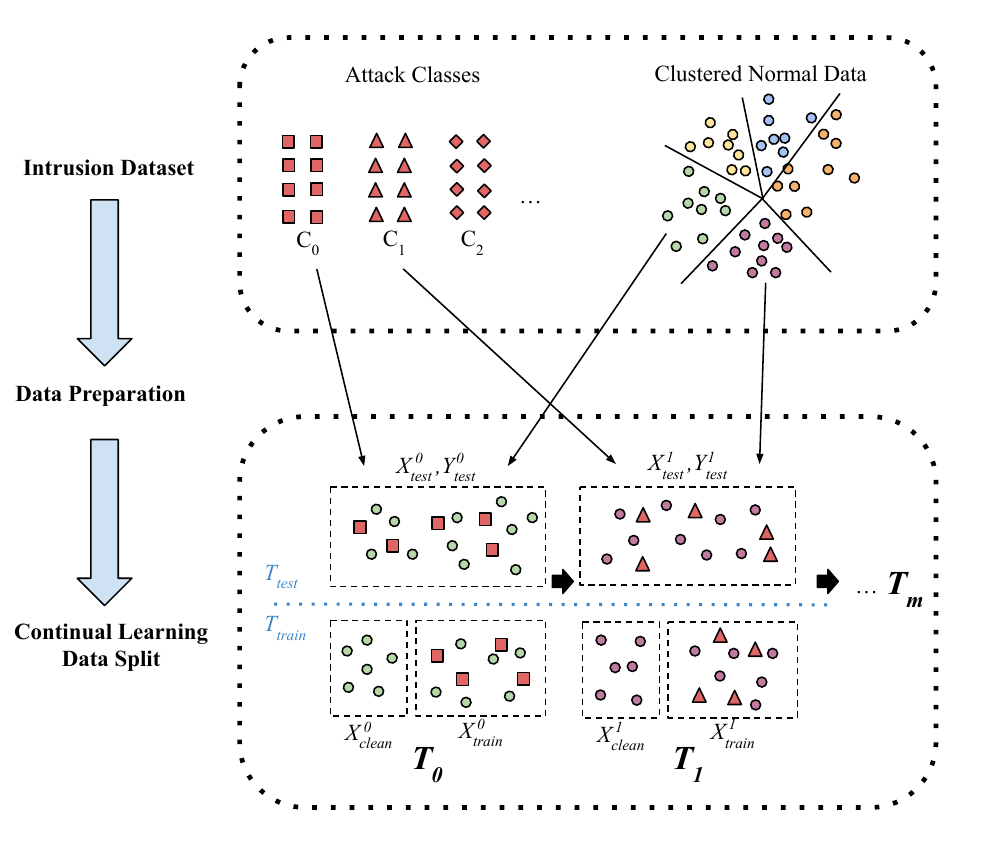}
    \caption{Overview of the CL data preparation process.}
    \label{fig:data-prep}
\end{figure}

To simulate a realistic, evolving IDS environment, we construct a sequence of tasks $T = \{T_1, T_2, \dots, T_m\}$, where each task introduces a distributional shift relative to the previous one. This process is illustrated in Fig.~\ref{fig:data-prep}. Each task $T_i$ contains three components:
\[
\{ X_{\text{clean}}^i,\; X_{\text{train}}^i,\; (X_{\text{test}}^i, Y_{\text{test}}^i) \} \in T_i.
\]

The intrusion dataset is initially divided into normal data and attack classes. The normal data are clustered to capture their internal structure, while the attack classes are distributed across tasks to simulate evolving adversarial behavior. Consequently, each task contains a distinct subset of attack types and a separate region of the normal-data manifold. For every task $T_i$, we further divide its data into (i) a small clean subset $X_{\text{clean}}^i$ and (ii) an unlabeled training stream $X_{\text{train}}^i$ containing both normal and attack samples. A fully labeled test set $\{X_{\text{test}}^i, Y_{\text{test}}^i\}$ is also created. This setup models a realistic IDS environment with evolving distributions, limited clean data, and unlabeled streaming traffic.

\textbf{Evolving Attack Data.} We simulate evolving attack behavior by distributing the dataset’s unique attack classes $|C|$ across the task sequence. Each task is assigned an approximately disjoint subset of $\tfrac{|C|}{m}$ attack types (e.g., $T_0$ may contain DDoS and Injection, while $T_1$ introduces Backdoor and Worms). Consequently, the model encounters previously unseen attack patterns at each stage.

\textbf{Evolving Normal Data.} To model gradual drift in benign behavior, we adopt the clustering procedure of Faber et al.~\cite{Faber_2024}, partitioning the aggregate normal data into $m$ groups using Minibatch K-Means~\cite{sculley2010web}. For task $T_i$, normal samples closest to the $i$-th cluster center are selected. A small portion is held out as the Reference Clean Set ($X_{\text{clean}}^i$), while the remainder is incorporated into the Unlabeled Stream ($X_{\text{train}}^i$). This setup produces evolving normal distributions that change in tandem with the evolving attacks.

\textbf{Train-Test Split.} Each task's training set is divided into:
\begin{enumerate}
    \item \textbf{Reference Clean Set} ($X_\text{clean}^i$): a small amount of clean data used to train the novelty detector;
    \item \textbf{Unlabeled Stream} ($X_{\text{train}}^i$): a larger, contaminated collection of normal and attack samples.
\end{enumerate}

This setup reflects a realistic IDS scenario in which obtaining a small amount of clean normal data is feasible, whereas labeling the entire deployment stream is costly. Consequently, $X_{\text{train}}^{i}$ is intentionally left unlabeled.

During training, \Design{} is trained on each task sequentially and only accesses $X_\text{clean}^i$ and $X_{\text{train}}^i$ from the current task. After completing task $T_i$, the model is evaluated on the test sets of \emph{all} tasks. Testing on earlier tasks ($T_j$ with $j < i$) measures \textit{catastrophic forgetting}, while testing on later tasks ($T_j$ with $j > i$) measures \textit{zero-day attack detection} performance.

For example, suppose:
\begin{itemize}
    \item $T_0$ contains attacks \textit{DDoS} and \textit{Injection}.
    \item $T_1$ contains attacks \textit{Backdoor} and \textit{Worm}.
    \item $T_2$ contains attacks \textit{Fuzzing} and \textit{Shellcode}.
\end{itemize}
After training on $T_1$, evaluation proceeds as follows:

\begin{itemize}
    \item $T_0$ (past): forgetting on DDoS/Injection.
    \item $T_1$ (current): performance on Backdoor/Worm.
    \item $T_2$ (future): zero-day detection on Fuzzing/Shellcode.
\end{itemize}

Because \Design{} learns tasks sequentially, performance on $T_0$ indicates how well earlier knowledge is retained, performance on $T_1$ reflects learning on the current task, and performance on $T_2$ reveals the model’s ability to identify previously unseen attacks.

\subsection{Framework Overview}

Fig.~\ref{figure:model_diagram} illustrates the end-to-end data flow of \Design{} during training and inference. We assume access only to a small trusted benign reference set (clean normal traffic) and an unlabeled stream that may contain both benign and malicious samples. Each incoming batch is processed by five components (A–E) that assign cluster-based pseudo-labels, maintain replay buffers, learn continual representations, and compute anomaly scores for zero-day and previously seen attacks.

\Design{} begins by updating the available normal data through  the \textbf{(B) normal-data memory update}. This module retains representative normal samples from past tasks and merges them with newly observed clean data, ensuring that the system maintains a stable reference of benign behavior over time. The aggregated normal data is then passed to the \textbf{(A) K-Means pseudo-labeler}, which leverages these clean samples to assign cluster-based pseudo-labels to the unlabeled training stream. Once pseudo-labels are assigned, the labeled training data is routed through the \textbf{(C) train-data memory update}, where past training samples—along with their pseudo-labels—are selectively stored to mitigate catastrophic forgetting in later stages.

Next, the pseudo-labeled training data enters the \textbf{(D)  Continual Feature Extractor} (CFE). The CFE is an autoencoder-based module that learns evolving embeddings of the data stream. It is optimized using a combination of metric-learning and reconstruction objectives, enabling the encoder to maintain separation between normal and anomalous patterns while adapting to shifting distributions. Both normal and inference-time samples are encoded through this module, producing the latent representations used for anomaly scoring.

Finally, encoded features are passed to the \textbf{(E) PCA-based novelty detector}. PCA is fit exclusively on encoded clean normal data, allowing it to model the expected subspace of benign behavior. During inference, each encoded sample is reconstructed by PCA, and its reconstruction error is used as a continuous anomaly score. A thresholding mechanism then converts this score into a binary decision—normal or attack.

Together, components (A–E) empower \Design{} to continuously adapt to evolving benign and malicious traffic, retain robust long term representations, and reliably detect attacks without requiring any labeled intrusion data.

\subsection{Detailed Methodology}
\textbf{A) K-Means Pseudo-Labeler.} After the continual learning data preparation phase, each task comprises three data splits: a small set of clean normal data $X_{\text{clean}}$, a large unlabeled training set containing both normal and attack samples $X_{\text{train}}$, and a fully labeled test set $(X_{\text{test}}, Y_{\text{test}})$. The K-Means pseudo-labeler leverages the limited supervision available in $X_{\text{clean}}$ to assign binary pseudo-labels $\hat{Y}_{\text{train}} \in \{0, 1\}$ to samples in $X_{\text{train}}$.

Pseudo-label assignment proceeds as follows. First, the samples in $X_{\text{train}}$ are clustered into $K$ clusters. Each sample in $X_{\text{clean}}$ is then associated with its nearest cluster center, and all clusters containing at least one clean sample are designated as normal clusters and assigned a pseudo-label of $\hat{Y}_{\text{train}} = 0$. All remaining clusters are assigned a pseudo-label of $\hat{Y}_{\text{train}} = 1$, corresponding to anomalous data. Finally, each sample in $X_{\text{train}}$ inherits the pseudo-label of its nearest cluster center. This procedure allows the small clean reference set to propagate supervision to the unlabeled stream.

We implement this pseudo-labeling strategy using MiniBatchKMeans~\cite{sculley2010web}. Although the proposed labeling mechanism is agnostic to the choice of clustering algorithm, MiniBatchKMeans is particularly well suited for large-scale intrusion datasets due to its computational efficiency.

\textbf{B) and C) Normal-Data and Train-Data Memory Update.} For both the normal data memory and the training data memory, we employ reservoir-based sampling~\cite{chaudhry2019tiny}. This strategy maintains a fixed-size memory buffer while processing a potentially unbounded data stream by ensuring that each incoming sample has an equal probability of being retained, independent of its arrival time. As a result, the memory buffer provides an unbiased and representative subset of past data, which is critical for mitigating catastrophic forgetting in continual learning settings. In practice, we find this sampling strategy to be as effective, if not more, than more complex memory management techniques. Section~\ref{Section:memory_anaylsis} presents a comparative study of different memory management approaches.

The \textbf{normal-data memory module} is applied \emph{prior to pseudo-labeling} to update the clean reference set $X_{\text{clean}}$, ensuring that it remains representative of benign behavior observed across past tasks. In contrast, the \textbf{training-data memory module} is applied \emph{after pseudo-labeling} to the combined set $(X_{\text{train}}, \hat{Y}_{\text{train}})$, allowing previously assigned pseudo-labeled samples to be retained and replayed during subsequent training. We find that preserving pseudo-labels from past tasks is critical for mitigating catastrophic forgetting, as they encode the model’s historical decision structure and provide continuity across tasks. Section~\ref{Section:ablation_study} highlights the performance degradation observed when pseudo-labels from past tasks are excluded from the memory module.

\textbf{D) Continual Feature Extractor (CFE).} CFE is an autoencoder (AE)-based model that leverages Multi-Layer Perceptron (MLP) as both the encoder and decoder. 
Our continual feature extractor loss ($L_{CFE}$) consists of a metric learning loss ($L_{M}$) and a reconstruction loss ($L_{R}$): 

\begin{equation}
    \label{equation:final_loss}
    L_{CFE} = L_{M} + L_R 
\end{equation}

$\triangleright$ \textbf{Metric Loss.} Using the previously described pseudo-labels, the final loss is computed using the widely adopted triplet margin loss \cite{schroff2015facenet}, defined as:
\begin{equation}
    L_{M} = \max\left( \Delta_{ap} - \Delta_{an} + m, 0 \right)
\end{equation} where $\Delta_{ap}$ represents the Euclidean distance between an anchor point and a positive point (a point of the same class), and $\Delta_{an}$ is the Euclidean distance between the anchor and a negative point (a point from a different class). The term $m > 0$ is a hyper-parameter that specifies the desired margin between positive and negative points.

Optimizing the metric loss structures the latent space according to Euclidean distance, which aligns with the geometry used by the PCA-based novelty detector. As noted in \cite{shlens2014tutorial}, under a mean squared error (i.e., $\ell_2$) loss, PCA provides the optimal low-dimensional representation of the data. By encouraging pseudo-normal samples to form compact clusters while pushing pseudo-anomalous samples farther apart, the triplet loss increases variance along directions orthogonal to the normal subspace. This, in turn, directly influences the PCA reconstruction objective, resulting in lower reconstruction error for normal samples and higher error for anomalous samples.

$\triangleright$ \textbf{Reconstruction Loss.} The reconstruction loss ensures that the features embedded by the encoder retain significant information from the original data. We found that this approach helps the PCA extractor better generalize to future and past tasks by forcing the model to learn encodings that align more closely with the original data. The reconstruction loss is the mean squared error (MSE) between the original input and the reconstructed output of the decoder. Let $h$ be the encoded feature of point $x$. Then $L_R$ is formulated as: 
\begin{equation}
L_R = MSE(\text{decoder}(h) , x)
\end{equation}

\textbf{E) PCA-based Novelty Detection.} Inspired by \cite{rios2022incdfm}, we create a PCA-based novelty detection approach. Let $\textbf{h}$ denote the output feature from our feature extractor. We utilize principal component analysis transformation $\textbf{T}$ to map the input feature from a high dimensional space to low dimensional space $\textbf{T} : \textbf{h} \rightarrow \textbf{l}$. We then utilize the feature reconstruction error ($FRE$) as the anomaly score: $FRE = \norm{\textbf{h} - \textbf{T}^{-1}(\textbf{l})} ^2$, where $\textbf{T}^{-1}$ is the inverse PCA transform. The anomaly score is therefore the reconstruction loss from this method. We train the PCA transformation on the subset of clean data $X^i_{\text{clean}}$ after it is encoded by the CFE. 

$\triangleright$ \textbf{Thresholding.} After computing the reconstruction error, each sample is classified as normal or attack by applying a threshold. To determine this threshold, we reserve a small validation subset from $X^i_{\text{clean}}$ (20\% in our experiments). We compute the mean and standard deviation of the reconstruction errors on this subset, and set the decision threshold to $\mu + 2\sigma$. Assuming an approximately Gaussian error distribution, this threshold corresponds to an expected false-positive rate of roughly 2.3\% on benign validation data.

\subsection{\Design{} Algorithm}
Algorithm 1 summarizes the training and evaluation procedure of \Design{} across a sequence of tasks. For each task, the framework receives an unlabeled training stream and a small set of clean normal samples. Clean normal data is first incorporated into a reservoir-based normal memory, which maintains a representative set of normal samples from past tasks. The current unlabeled training data are then clustered using K-Means, and pseudo-labels are assigned by identifying clusters that overlap with stored clean normal data. These pseudo-labeled samples are stored in a train memory buffer and replayed during training to mitigate forgetting. The continual feature extractor is trained on the replayed data using a combination of metric-learning and reconstruction objectives, enabling the model to maintain discriminative representations as the data distribution evolves. After training, clean normal samples are encoded and used to fit a PCA-based novelty detector, with anomaly scores computed as reconstruction error. A statistical threshold is determined from clean validation data, and the resulting detector is evaluated on the union of test sets from all tasks. This process is repeated after each task, enabling continual adaptation while preserving detection performance on previously observed data.

\begin{algorithm}[t]
\caption{ACORN-IDS Training and Evaluation over Tasks}
\label{alg:train_test}
\begin{algorithmic}[1]
\Require Tasks $T=\{T_1,\dots,T_{m}\}$; 
each $T_i$ provides $(X^i_{\text{train}}, X^i_{\text{clean}}, X^i_{\text{test}}, Y^i_{\text{test}})$
\Require NM, TM: reservoir memories; 
CFE: feature extractor; 
PL: K-Means pseudo-labeler; 
ND: PCA detector
\State Initialize normal memory $\mathcal{M}_N \leftarrow \emptyset$
\State Initialize train memory $\mathcal{M}_T \leftarrow \emptyset$

\For{$i = 1$ to $|T|$}
    \State Obtain $(X^i_{\text{train}}, X^i_{\text{clean}}, X^i_{\text{test}}, Y^i_{\text{test}}) \in T_i$

    \State Add Memory ${X}^i_{\text{clean}} = \mathcal{M}_N + {X}^i_{\text{clean}} $
    \State Update $\mathcal{M}_N \leftarrow \textsc{Reservoir}(\mathcal{M}_N, X^i_{\text{clean}})$

    \State Fit PL on $X^i_{\text{train}}$; obtain $K$ clusters
    \State Assign pseudo-labels $\hat{Y}^i_{\text{train}}$ using $X^i_{\text{clean}}$
     \State Add Memory $(X^i_{\text{train}},\hat{Y}^i_{\text{train}}) = \mathcal{M}_T + (X^i_{\text{train}},\hat{Y}^i_{\text{train}})$
    \State Update $\mathcal{M}_T \leftarrow 
    \textsc{Reservoir}(\mathcal{M}_T,(X^i_{\text{train}},\hat{Y}^i_{\text{train}}))$

    \State Train CFE on $(X^i_{\text{train}},\hat{Y}^i_{\text{train}})$
    \State Encode clean normal data $H_c \leftarrow \text{CFE}(X^i_{\text{clean}})$
    \State Fit PCA on $H_c$; use reconstruction error as score

    \State Compute $\mu,\sigma$ from validation clean data
    \State Set threshold $\tau_i \leftarrow \mu + 2\sigma$

    \State $X_{\text{test}}, Y_{\text{test}} \leftarrow \bigcup_{j=1}^{|T|} X^j_{\text{test}}, Y^j_{\text{test}}$
    \State $S_{\text{test}} \leftarrow \text{ND}(\text{CFE}(X_{\text{test}}))$
    \State $\hat{Y}_{\text{pred}} \leftarrow 1[S_{\text{test}} > \tau_i]$

    \State Compute evaluation metrics
\EndFor
\end{algorithmic}
\end{algorithm}

\section{Experimental Setup}
\label{experimental}
\textbf{Preprocessing.} All datasets are processed using a unified preprocessing pipeline. Categorical features are encoded via one-hot encoding, while numerical features are standardized using the \texttt{RobustScaler}~\cite{scikit-learn} to mitigate sensitivity to outliers. Subsequently, each feature is min--max normalized to the range $[0,1]$ by subtracting the feature-wise minimum and dividing by the feature-wise maximum. This normalization ensures consistent feature scaling across datasets and contributes to more stable model training.

\begin{table}[]
    \caption{Selected Intrusion Datasets}
    \centering
    \label{tab:datasets}
    \resizebox{\columnwidth}{!}{
    \begin{tabular}{|c|c|c|c|c|}
    \hline
    Dataset    & Size      & Normal Data & Attack Data & Attack Types \\ 
    \hline
    X-IIoTID \cite{al2021x}   & 820,502   & 421,417     & 399,417     & 18           \\
    \hline
    WUSTL-IIoT \cite{zolanvari2021wustl} & 1,194,464 & 1,107,448   & 87,016      & 4       \\
    \hline
    CICIDS2017 \cite{Sharafaldin2018TowardGA} & 2,830,743 & 2,273,097 & 557,646 & 15 \\
    \hline
    CICIDS2018 \cite{cicids2018} & 16,232,943 & 13,484,708& 2,748,235 & 15 \\
    \hline
    UNSW-NB15 \cite{moustafa2015unsw}
 & 257,673 & 164,673 & 93,000 & 10 \\
    \hline
    \end{tabular}}
\end{table}
\textbf{Datasets.} Table~\ref{tab:datasets} provides a detailed overview of the intrusion datasets used in our study. We select these datasets based on their recency and realism, reflecting modern attack patterns and network/IIoT environments. For IIoT intrusion, we employ X-IIoTID~\cite{al2021x} and WUSTL-IIoT~\cite{zolanvari2021wustl}, which capture realistic IIoT traffic and attack scenarios. We also include widely used network intrusion datasets CICIDS2017~\cite{Sharafaldin2018TowardGA}, CICIDS2018~\cite{cicids2018}, and UNSW-NB15~\cite{moustafa2015unsw}, which provide diverse attack types and reflect contemporary network conditions. For X-IIoTID, CICIDS2017, CICIDS2018, and UNSW-NB15, the data are split across five tasks, with each task containing two to four attack types. For WUSTL-IIoT, the data are split across four tasks, with each task containing a single attack type. 

\textbf{Continual Learning Setup.} We evaluate all methods under two realistic continual learning scenarios:  
(1) \textbf{Evolving Attacks (EA)}, where only the attack distribution changes while the normal data remains stationary; and  
(2) \textbf{Evolving Normal and Attack Distributions (ENA)}, where both benign traffic and attack patterns evolve over time.  
In the EA scenario the attack classes evolve throughout tasks, whereas in the ENA scenario the normal data distribution also evolves through evolving cluster centers. The specific procedure for generating these distribution shifts is described in Section~\ref{CL_data_prep}.

\textbf{Hyper-Parameters.} We utilize \textit{elbow method} \cite{han2011data} for determining the number of K-Means Pseudo-labeler clusters $K$. For the triplet margin loss $m$ we use 0.2 after careful experimentation. For the AE modules of \Design{}, we use 4-layer MLP with [256, 512, 256] neurons in the hidden layers. We train it using Adam optimizer with a learning rate of 0.001. For PCA, we select the number of principal components using the explained variance criterion, retaining components that collectively account for 95\% of the total variance \cite{rios2022incdfm}. The Normal-Data Memory and Train-Data Memory both hold a maximum of 10K data points. For all experiments, we perform three independent runs and report the mean performance.

\textbf{State-of-the-art Baselines.} We compare \Design{} against a broad set of continual learning and novelty detection approaches. These baselines fall into three categories:

\textbf{(1) Unsupervised Continual Learning (UCL) Methods}
\begin{itemize}
    \item \textbf{ADCN} \cite{ashfahani2023unsupervised}: An autonomous deep clustering network that incrementally updates cluster structures and representations from an unsupervised data stream.
    \item \textbf{LwF} \cite{lwf2019Li}: An autoencoder combined with K-Means clustering and Learning without Forgetting  to mitigate catastrophic forgetting during sequential training.
\end{itemize}

ADCN and LwF require a small amount of labeled normal and attack data to classify samples, unlike \Design{}.

\textbf{(2) Continual Novelty Detection (CND) Method}
\begin{itemize}
    \item \textbf{VLAD} \cite{FABER_VLAD}: A variational autoencoder–based lifelong learning framework that incrementally models normal data distributions for anomaly detection.

\end{itemize}

\textbf{(3) Novelty Detection (ND) Methods}
\begin{itemize}
    \item \textbf{SLAD} \cite{xu2023fascinating}: Deep anomaly detection with scale learning for robust feature extraction.
    \item \textbf{ICL} \cite{shenkar2022anomaly}: Internal contrastive learning to enhance separability between normal and anomalous representations.
    \item \textbf{DIF} \cite{xu2023deep}: Deep Isolation Forest, an end-to-end differentiable extension of Isolation Forest.
    \item \textbf{AE} \cite{Torabi2023AutoencoderIDS}: A standard autoencoder trained solely on normal data.
    \item \textbf{PCA} \cite{rios2022incdfm}: Principal Component Analysis used for anomaly scoring based on reconstruction error.
    \item \textbf{OCSVM} \cite{scikit-learn}: One-Class SVM trained to model the boundary of normal data.
    \item \textbf{IF} \cite{scikit-learn}: Isolation Forest for unsupervised anomaly detection.
    \item \textbf{LOF} \cite{scikit-learn}: Local Outlier Factor based on density deviation.
\end{itemize}

The ND methods, including VLAD, are trained exclusively on normal data. Accordingly, in our two evaluation scenarios, these models experience a distribution shift only in the \textit{ENA} scenario. Therefore in the ENA scenario, we augment each ND baseline with the same reservoir memory mechanism used in \Design{}. We refit the detector after each task using the current task’s clean set augmented by the reservoir normal-memory. For all baselines that produce an anomaly score requiring thresholding (PCA, AE, DIF, ICL, SLAD, VLAD), we apply the same threshold-selection procedure as \Design{} to maintain consistency across methods.

\textbf{Evaluation Metrics.} To evaluate model performance, we primarily report the $F_{1}$ score. 
Since intrusion datasets are inherently imbalanced, $F_{1}$ provides a more reliable indicator of detection performance than accuracy. 
We assess performance at the end of each training task on all test sets, producing a matrix of $F_{1}$ scores $R_{ij}$, where $i$ is the current training task and $j$ is the testing task. 
From this matrix, we derive three widely used continual learning metrics \cite{diaz2018don}:

\begin{itemize}
    \item \textbf{Average $F_{1}$ (AVG):} 
    \[
    \text{AVG}_{F_1} = \frac{\sum_{i=1}^m R_{mi}}{m}
    \]
    Represents the average performance on all test tasks after the \textit{last} training task, measuring performance on \textit{seen attacks}.
    
    \item \textbf{Forward Transfer (FwdTrans):} 
    \[
    \text{FwdTrans}_{F_1} = \frac{\sum_{j > i} R_{ij}}{\tfrac{m(m-1)}{2}}
    \]
    Captures the average performance on \textit{future tasks}, simulating the model’s ability to generalize to zero-day attacks.
    
    \item \textbf{Backward Transfer (BwdTrans):} 
    \[
    \text{BwdTrans}_{F_1} = \frac{\sum_{i}^m R_{mi} - R_{ii}}{\tfrac{m(m-1)}{2}}
    \]
    Measures the change in performance on \textit{past tasks} after subsequent training. Negative values indicate catastrophic forgetting, while positive values suggest that learning new tasks improved past performance.
\end{itemize}

In summary, AVG reflects performance on seen attacks, FwdTrans measures generalization to unseen (zero-day) attacks, and BwdTrans quantifies forgetting. Higher positive values across all metrics indicate stronger performance. 

In the EA scenario, however, all ND (including VLAD) baselines are trained exclusively on normal data, and the normal data distribution remains stationary by construction. As a result, these methods do not experience a training-time distribution shift across tasks, since neither their training data nor their learned notion of normality changes over time. Consequently, task ordering has no effect on their learned models, rendering forward and backward transfer metrics ill-defined in this setting. For this reason, we report only the AVG $F_1$ score for ND methods under the EA scenario, which captures their steady-state detection performance against evolving attacks.

\textbf{Threshold-free Evaluation.} We also report the threshold-free metric Precision-Recall Area Under the Curve (PR-AUC) \cite{praucDavid06}. Since \Design{} requires selecting a threshold, PR-AUC allows us to assess model performance independently of the threshold. We choose PR-AUC over Receiver Operating Characteristic Area Under the Curve (ROC-AUC) because ROC-AUC can give misleadingly high results in the presence of class imbalance \cite{praucDavid06}.

\section{Results}
\subsection{Evolving Attack (EA) Scenario}

\begin{figure*}[t]
  \centering
  \includegraphics[width=\linewidth]{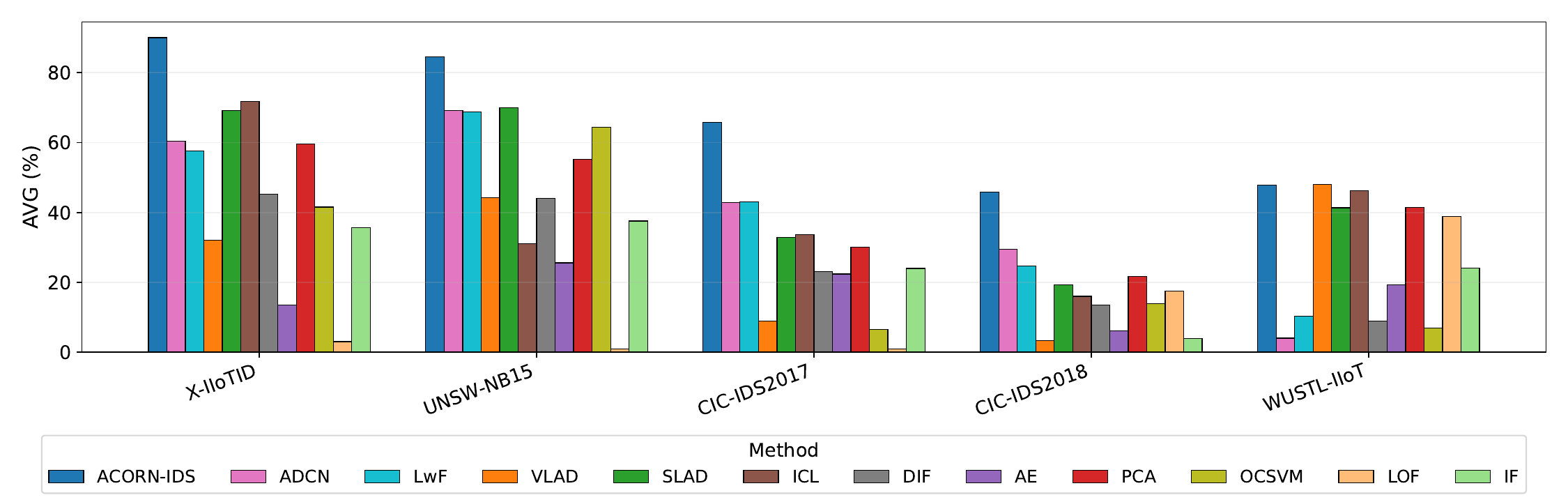}
  \caption{AVG score of \Design{} compared with all baselines under the \textit{EA} scenario across all datasets.}
  \label{fig:novelty_detectors}
\end{figure*}
\begin{table*}
\caption{BwdTrans and FwdTrans of \Design{} and all baselines on EA scenario.}
\label{tab:ea_bwt_fwt}
\begin{subtable}{0.48\textwidth}
    \centering
    \caption{Backward Transfer (\%)}
    \resizebox{\columnwidth}{!}{
    \begin{tabular}{lrrrrr|r}
        \toprule
        Dataset & XIIoT & UNSW & CICIDS17 & CICIDS18 & WUSTL & Avg $\pm$ Std \\
        Strategy &  &  &  &  &  &  \\
        \midrule
ACORN-IDS & \textbf{-0.49} & -2.12 & -2.41 & \textbf{-0.47} & -3.86 & -1.9 $\pm$ 1.4 \\
ADCN     & -3.72 & \textbf{-0.29} & \textbf{5.99} & -2.08 & \textbf{-1.95} & -0.4 $\pm$ 3.8 \\
LwF      & -0.77 & 4.14 & -3.45 & -0.74 & 2.96 & \textbf{0.4} $\pm$ 3.1 \\
        \bottomrule
    \end{tabular}
    }
\end{subtable}
\hfill
\begin{subtable}{0.48\textwidth}
    \centering
    \caption{Forward Transfer (\%)}
    \resizebox{\columnwidth}{!}{
    \begin{tabular}{lrrrrr|r}
        \toprule
        Dataset & XIIoT & UNSW & CICIDS17 & CICIDS18 & WUSTL & Avg $\pm$ Std \\
        Strategy &  &  &  &  &  &  \\
        \midrule
ACORN-IDS & \textbf{68.55} & \textbf{72.52} & 30.09 & 38.22 & \textbf{53.69} & \textbf{52.6} $\pm$ 18.5 \\
ADCN     & 47.80 & 54.03 & \textbf{35.37} & 28.61 & 0.00 & 33.2 $\pm$ 21.1 \\
LwF      & 61.16 & 66.18 & 31.21 & \textbf{43.07} & 23.20 & 45.0 $\pm$ 18.6 \\
        \bottomrule
    \end{tabular}
    }
\end{subtable}

\end{table*}

Fig.~\ref{fig:novelty_detectors} summarizes average $F_{1}$ (AVG) under the EA scenario, where the attack distribution evolves across tasks while the normal distribution remains stationary. In this setting, the strongest performance requires both (i) maintaining a stable notion of ``normal'' and (ii) adapting representations as new attack types appear in the unlabeled stream. Across all datasets, \Design{} achieves the highest AVG, and it exceeds the best competing novelty detector (SLAD) by an average 43.53\% increase in AVG. This gain highlights the effectiveness of \Design{} in leveraging the evolving unlabeled data stream to improve continual adaptation. Specifically, whereas traditional novelty detectors such as SLAD and ICL remain fixed after the initial task, \Design{} leverages the evolving unlabeled data stream to continually refine its latent representation. By applying a triplet margin loss on pseudo labeled clusters, the encoder learns to separate potential anomalies from the normal feature space. This structured representation provides increasingly discriminative inputs to the PCA based scoring module as the environment evolves, whereas static methods cannot exploit the additional data.

To quantify continual learning behavior, Table~\ref{tab:ea_bwt_fwt} reports backward and forward transfer for methods that continuously update their models across the task stream. \Design{} attains the highest average forward transfer (53.88\%), outperforming ADCN (33.16\%) and LwF (44.96\%), which suggests stronger generalization to future, unseen attacks. At the same time, \Design{} maintains near zero BWT overall, indicating limited forgetting while incorporating new attack information. Notably, some baselines exhibit positive BWT on certain datasets (e.g., ADCN on CICIDS2017 and LwF on UNSW/WUSTL), which can occur when later updates reshape feature space in a way that incidentally improves earlier-task decision boundaries. However, these positive BWT cases do not translate into the best overall detection performance: they are inconsistent across datasets and often coincide with weaker AVG and/or FWT. 

Overall, the EA results highlight a favorable stability–plasticity trade-off for \Design{}: it achieves the highest AVG performance while also exhibiting strong forward transfer.

\subsection{Evolving Normal and Attack Distributions (ENA)}

\begin{figure*}[t]
  \centering
  \includegraphics[width=\linewidth]{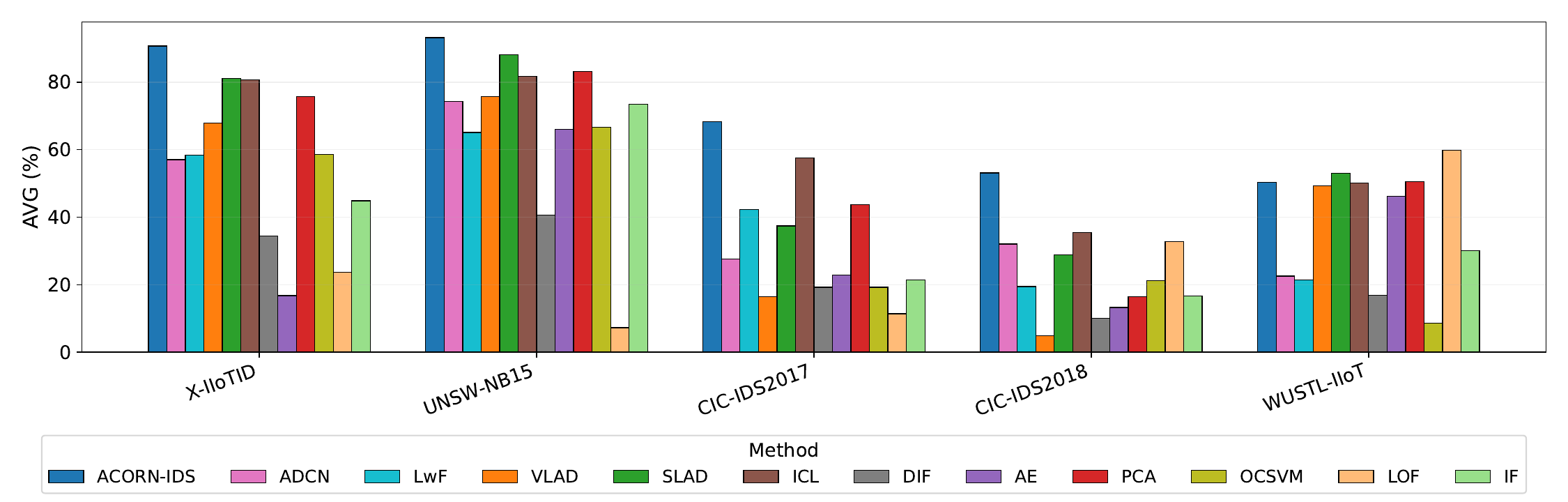}
  \caption{AVG score of \Design{} compared with all baselines under the \textit{ENA} scenario across all datasets.}
  \label{fig:ena_avg}
\end{figure*}

\begin{table}[t]
\centering
\caption{BwdTrans and FwdTrans of \Design{} and all baselines on ENA scenario.}
\label{tab:metrics}

\begin{subtable}{0.48\textwidth}
    \centering
    \caption{Backward Transfer (\%)}
    \resizebox{\columnwidth}{!}{
    \begin{tabular}{lrrrrr|r}
        \toprule
        Dataset & XIIoT & UNSW & CICIDS17 & CICIDS18 & WUSTL & Avg $\pm$ Std \\
        Strategy &  &  &  &  &  &  \\
        \midrule
ACORN-IDS & -1.92 & 0.09 & 1.73 & -1.55 & -1.97 & -0.7 $\pm$ 1.6 \\
ADCN & -3.27 & 2.11 & -5.17 & -4.45 & \textbf{19.74} & 1.8 $\pm$ 10.4 \\
LwF & \textbf{0.14} & 3.69 & 7.30 & -1.35 & -7.91 & 0.4 $\pm$ 5.7 \\
VLAD & -27.22 & -24.12 & -4.17 & -5.46 & -6.74 & -13.5 $\pm$ 11.2 \\
SLAD & -1.76 & -0.19 & -14.82 & -0.15 & -1.22 & -3.6 $\pm$ 6.3 \\
ICL & -2.26 & \textbf{6.26} & -10.19 & 0.96 & 1.39 & -0.8 $\pm$ 6.1 \\
DIF & -15.51 & -24.76 & -10.87 & -0.47 & 2.06 & -9.9 $\pm$ 11.0 \\
AE & -0.93 & 3.41 & -7.07 & 1.17 & 0.93 & -0.5 $\pm$ 4.0 \\
PCA & -6.07 & -0.72 & -21.64 & -3.74 & -0.55 & -6.5 $\pm$ 8.7 \\
IF & -3.62 & -5.65 & -10.01 & \textbf{20.23} & 7.55 & 1.7 $\pm$ 12.2 \\
LOF & -3.23 & -38.94 & -7.62 & -1.61 & -28.80 & -16.0 $\pm$ 16.8 \\
OCSVM & -0.44 & 2.38 & \textbf{7.84} & -1.72 & 0.88 & \textbf{1.8 }$\pm$ 3.7 \\
        \bottomrule
    \end{tabular}
    }
\end{subtable}
\hfill
\begin{subtable}{0.48\textwidth}
    \centering
    \caption{Forward Transfer (\%)}
    \resizebox{\columnwidth}{!}{
    \begin{tabular}{lrrrrr|r}
        \toprule
        Dataset & XIIoT & UNSW & CICIDS17 & CICIDS18 & WUSTL & Avg $\pm$ Std \\
        Strategy &  &  &  &  &  &  \\
        \midrule
ACORN-IDS & 62.97 & \textbf{67.25} & 28.60 & \textbf{45.61} & 81.61 & 57.2 $\pm$ 20.5 \\
ADCN & 32.30 & 57.49 & 16.06 & 28.00 & 0.00 & 26.8 $\pm$ 21.3 \\
LwF & 65.69 & 59.41 & 35.94 & 34.47 & 42.00 & 47.5 $\pm$ 14.2 \\
VLAD & 46.86 & 52.86 & 32.45 & 5.23 & 70.17 & 41.5 $\pm$ 24.4 \\
SLAD & 69.65 & 49.75 & \textbf{52.54} & 39.38 & \textbf{82.07} & \textbf{58.7} $\pm$ 17.0 \\
ICL & 49.94 & 40.17 & 46.84 & 30.66 & 56.78 & 44.9 $\pm$ 9.9 \\
DIF & 16.53 & 14.64 & 34.33 & 5.92 & 28.22 & 19.9 $\pm$ 11.3 \\
AE & 37.46 & 55.12 & 44.71 & 14.19 & 79.32 & 46.2 $\pm$ 23.9 \\
PCA & \textbf{74.99} & 54.73 & 50.32 & 17.05 & 80.72 & 55.6 $\pm$ 25.1 \\
IF & 54.72 & 57.35 & 42.44 & 7.34 & 48.35 & 42.0 $\pm$ 20.2 \\
LOF & 15.81 & 26.14 & 8.77 & 28.13 & 71.29 & 30.0 $\pm$ 24.4 \\
OCSVM & 68.24 & 50.74 & 14.79 & 28.29 & 23.26 & 37.1 $\pm$ 21.9 \\
        \bottomrule
    \end{tabular}
    }
\end{subtable}

\end{table}

In the ENA scenario, both the benign traffic and attack distributions evolve across tasks, creating a setting where models must simultaneously track changes in “normal” behavior and adapt to emerging attacks. Figure~\ref{fig:ena_avg} reports the AVG $F_1$ score for all methods under this configuration. Across datasets, \Design{} achieves the highest or near-highest AVG score, indicating consistently strong performance on seen attacks even as both normal and malicious traffic patterns change over time. In this scenario, the most competitive baseline is ICL; however \Design{} achieves a 18.46\% higher AVG. Similar to the EA scenario, the performance gap between \Design{} and ND methods like ICL or SLAD highlights the benefits of incorporating the unlabeled data stream. 

Table~\ref{tab:metrics} summarizes the corresponding per-dataset BWT and FWT values. BWT results further demonstrate the resilience of \Design{} to catastrophic forgetting. Its BWT values remain close to zero on all datasets, reflecting minimal forgetting of past tasks. In contrast, most baselines exhibit substantially negative BWT (e.g., DIF and LOF on UNSW), confirming that they struggle to retain performance as the data distribution shifts. A few methods show large positive BWT on specific datasets—for example, ADCN and IF on WUSTL-IIoT—which suggests that later tasks can sometimes improve the performance of past tasks; however, this behavior is inconsistent and often accompanied by poorer AVG or FWT on other datasets.  It is worth noting that all ND (non-continual) baselines are enhanced with the same Normal-Data Memory as \Design{}, however \Design{} still exhibits superior BWT. This highlights the benefit of the dual-memory architecture of \Design{}, The Normal-Data Memory ensures the PCA detector retains a historical anchor of benign behavior, while the Train-Data Memory provides the CFE with the stability needed to avoid catastrophic forgetting during the representation learning phase.

The forward transfer results in Table~\ref{tab:metrics} show that \Design{} attains competitive FWT across all datasets and is often among the top-performing methods. Notably, ACORN-IDS outperforms the UCL baselines (ADCN and LwF) on FWT for most datasets, underscoring the benefit of explicitly modeling normal data via continual ND.  
Relative to ND baselines, \Design{} maintains comparable FWT while delivering substantially higher AVG, indicating that leveraging the unlabeled stream improves overall detection without compromising transfer to future attacks. This behavior is supported by the reconstruction objective and PCA-based scoring, which encourage consistent structure in the learned representation as the stream evolves.

Overall, the ENA results show that \Design{} maintains high detection performance, strong forward transfer, and low forgetting even when both normal and attack traffic evolve. This robustness under simultaneous distribution shifts is critical for realistic IDS deployments, where benign behavior and adversarial strategies rarely change in isolation.

\subsection{Threshold Free Evaluation}

\begin{figure}[t]
  \centering
  \includegraphics[width=0.9\linewidth]{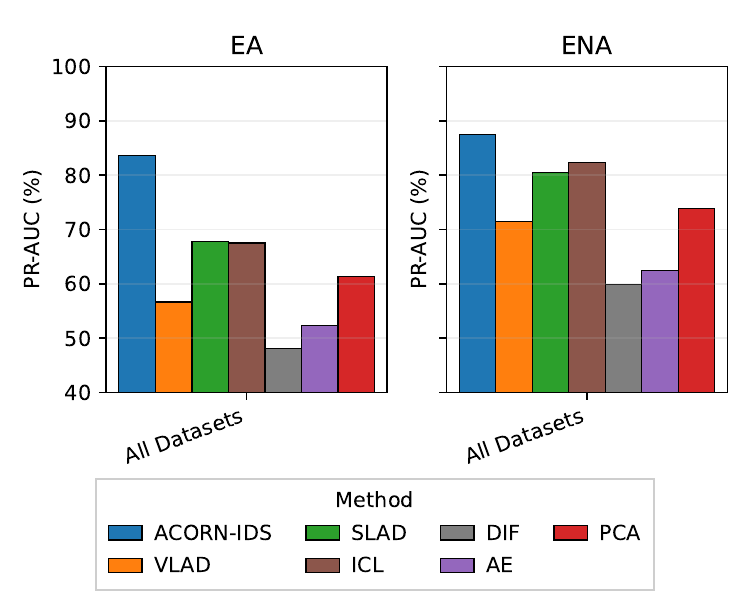}
  \caption{Threshold-free performance (PR-AUC) of \Design{} and relevant baselines, averaged across all datasets.}
  \label{fig:ea_threshold_free}
\end{figure}

\begin{figure*}[t]
  \centering
  \includegraphics[width=\linewidth]{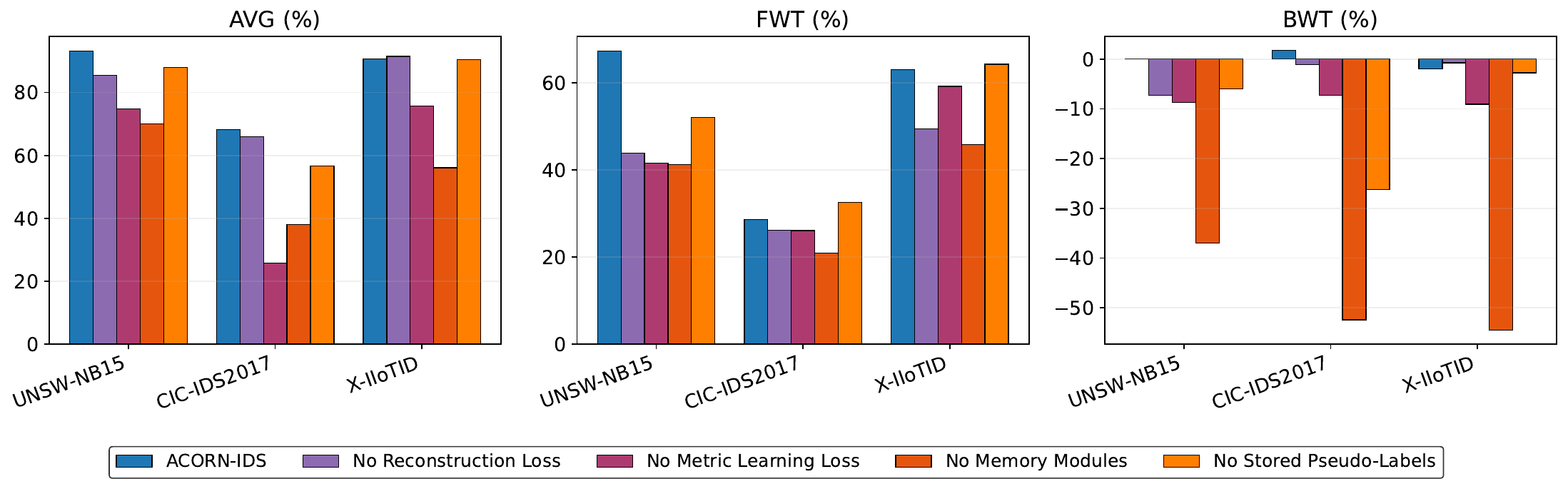}
  \caption{Ablation study of \Design{} in the ENA scenario}
  \label{fig:ablation_study}
\end{figure*}

\begin{figure*}[t]
  \centering
  \includegraphics[width=\linewidth]{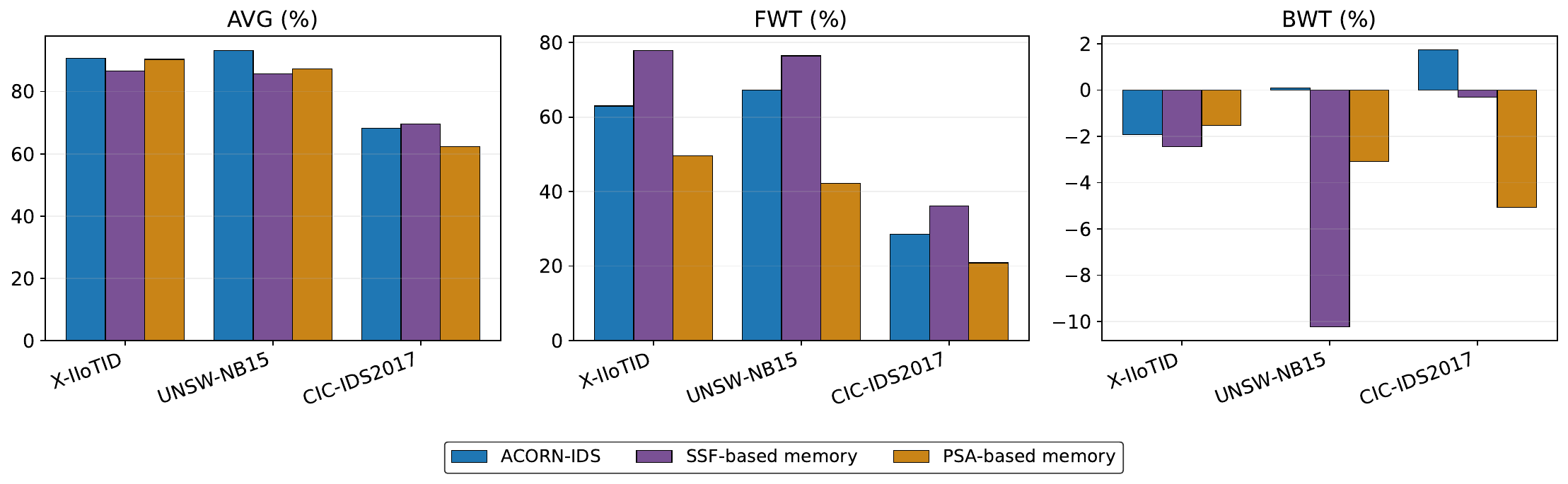}
  \caption{Memory type analysis of \Design{} in the ENA scenario}
  \label{fig:memory_study}
\end{figure*}

\begin{table*}[t]
\centering
\caption{Average inference time (in ms) per test sample}
\label{tab:overhead}
\begin{tabular}{lccccccc}
\toprule
\textbf{Strategy} & \Design{} & ADCN & LwF & SLAD & ICL & DIF & PCA \\
\midrule
\textbf{Inference Time (ms)}
& \underline{0.005}
& 0.406
& 0.067
& 0.924
& 0.084
& 1.053
& \textbf{0.002} \\
\bottomrule
\end{tabular}
\end{table*}

To evaluate intrusion detection quality independently of any thresholding strategy, we report threshold-free Precision–Recall AUC (PR-AUC), shown in Fig.~\ref{fig:ea_threshold_free}. For a fair comparison, we consider only methods that produce continuous anomaly scores and therefore support an equivalent thresholding procedure. 

Across both the EA and ENA scenarios, \Design{} consistently achieves the highest PR-AUC when averaged over all datasets, closely mirroring the trends observed in the thresholded $F_{1}$ results. This agreement indicates that the performance gains of \Design{} are not an artifact of threshold selection, but rather reflect superior anomaly scoring. 

The higher PR-AUC achieved by \Design{} relative to standard Autoencoders (AE) and PCA indicates that the self-supervised triplet loss effectively increases the separation between normal and anomalous samples in the latent space. In a standard AE, anomalies might still be reconstructed with relatively low error if they share local features with normal data. In contrast, the \Design{} metric learning objective explicitly penalizes the proximity of pseudo-anomalies to the normal cluster. This results in a more dramatic spike in reconstruction error at inference time, leading to a clearer separation in the Precision-Recall curve regardless of where the final binary decision boundary is drawn.

Overall, the close correspondence between pre-threshold and post-threshold performance indicates that \Design{} learns feature representations that achieve a clear separation between normal and anomalous samples.

\subsection{Ablation Study}
\label{Section:ablation_study}
For our ablation study, we evaluate the contribution of each component of \Design{} under the ENA scenario using three datasets: CIC-IDS2017, UNSW-NB15, and X-IIoTID. As shown in Fig.~\ref{fig:ablation_study}, we assess performance after separately removing (i) the metric learning loss, (ii) the reconstruction loss, (iii) the train/normal-data memory modules, and (iv) the stored pseudo-labels used in the training memory buffer. In case (iv), instead of storing full $(X, Y)$ pairs for replay, the buffer stores only the raw inputs $X$, and pseudo-labels are recomputed using the current task’s pseudo-labeler when samples are reloaded.

The results show that both the metric learning and reconstruction objectives play a substantial role in improving detection accuracy and mitigating forgetting. This observation aligns with prior work demonstrating that unsupervised objectives promote smoother and more stable loss landscapes in continual learning \cite{madaan2022representational}. Moreover, retaining past-task pseudo-labels is essential: removing them leads to severe degradation, indicating that pseudo-labels encode valuable information about the model’s historical decision boundaries and help preserve its prior knowledge. Finally, the memory buffer provides significant help to alleviate catastrophic forgetting, and the full \Design{} configuration—combining all components—achieves the most reliable and consistent performance across datasets.

\subsection{Memory Type Analysis}
\label{Section:memory_anaylsis}
We also analyze alternative memory designs and motivate our choice of \emph{reservoir sampling} as the memory mechanism for \Design{}. We compare three memory strategies under the ENA scenario: (i) reservoir sampling \cite{chaudhry2019tiny}, used in \Design{}; (ii) a Strategic Selection and Forgetting (SSF)-based memory strategy \cite{zhang2024ssf}; and (iii) a Part-and-Select Algorithm (PSA)-based memory strategy derived from \cite{yu2023scale}. All methods operate under an identical memory budget and are evaluated using Average $F_{1}$ (AVG), forward transfer (FWT), and backward transfer (BWT), as shown in Fig.~\ref{fig:memory_study}.

Overall, \Design{} equipped with \emph{reservoir} memory achieves the strongest and most consistent performance, yielding the highest AVG and the most stable BWT across datasets. SSF-based memory exhibits noticeably higher FWT than reservoir sampling, suggesting that its strategic selection mechanism may promote stronger generalization to future tasks. However, this gain in forward transfer comes at the cost of substantially worse BWT, indicating increased forgetting of past tasks, likely due to poor selection of data removal. In contrast, PSA-based memory underperforms across most metrics, reflecting the difficulty of reliably estimating sample importance in an unsupervised, pseudo-label-driven setting.

Taken together, these results indicate that reservoir sampling provides the best overall trade-off between detection accuracy, knowledge retention, and forward generalization. While more selective memory strategies can improve specific metrics such as FWT, they tend to sacrifice stability on past tasks. This analysis highlights the critical role of memory design in continual novelty detection and suggests that developing memory mechanisms that jointly optimize forward transfer and forgetting remains an important direction for future work.

\subsection{Overhead Analysis}
In real-world intrusion detection systems, computational efficiency is critical, as IDS must process high-volume network traffic in real time. Table~\ref{tab:overhead} compares the inference overhead of \Design{} with ADCN, LwF, SLAD, ICL, DIF, and PCA. \Design{} achieves the fastest inference time among continual learning methods and novelty detectors, second only to PCA. This efficiency arises because \Design{} avoids clustering-based classification used by LwF and ADCN and relies on PCA-based reconstruction, which is computationally lightweight. The difference between \Design{} and PCA is minimal—only 0.003 milliseconds slower due to the additional but lightweight encoding step. Considering that the average median flow duration across datasets is 27.77 milliseconds, the overhead introduced by \Design{} is negligible for real-time traffic processing.

\section{Conclusion}
\label{conclusion}
This work introduced \Design{}, an adaptive continual novelty detection framework for intrusion detection that addresses two key challenges in real-world IDS deployment: non-stationary traffic distributions and the lack of labeled attack data. \Design{} combines a dual-memory architecture with a continual feature extractor trained using reconstruction and metric-learning objectives, and employs a PCA-based reconstruction module for anomaly scoring.
This design enables continuous adaptation to distributional shifts in both benign and malicious traffic without relying on attack labels. Extensive evaluation on five realistic intrusion datasets under two continual learning scenarios demonstrates that \Design{} substantially outperforms state-of-the-art unsupervised continual learning and novelty detection baselines. In particular, it achieves significant gains in F1-score and zero-day attack detection, exhibits near-zero forgetting, and maintains minimal inference overhead suitable for real-time deployment. Overall, these results highlight continual novelty detection as a practical, label-efficient paradigm for building robust and adaptive intrusion detection systems capable of operating effectively in dynamic, real-world environments.

\section*{Acknowledgments}
This work has been funded in part by NSF, with award numbers \#1826967, \#1911095, \#2003279, \#2052809, \#2100237, \#2112167, \#2112665, and in part by PRISM and CoCoSys, centers in JUMP 2.0, an SRC program sponsored by DARPA.

\bibliographystyle{ieeetr}
\bibliography{bibfile}

@inproceedings{xu2023fascinating,
  title={Fascinating supervisory signals and where to find them: Deep anomaly detection with scale learning},
  author={Xu, Hongzuo and Wang, Yijie and Wei, Juhui and Jian, Songlei and Li, Yizhou and Liu, Ning},
  booktitle={International Conference on Machine Learning},
  pages={38655--38673},
  year={2023},
  organization={PMLR}
}

@inproceedings{shenkar2022anomaly,
  title={Anomaly detection for tabular data with internal contrastive learning},
  author={Shenkar, Tom and Wolf, Lior},
  booktitle={International conference on learning representations},
  year={2022}
}

@article{xu2023deep,
  title={Deep isolation forest for anomaly detection},
  author={Xu, Hongzuo and Pang, Guansong and Wang, Yijie and Wang, Yongjun},
  journal={IEEE Transactions on Knowledge and Data Engineering},
  volume={35},
  number={12},
  pages={12591--12604},
  year={2023},
  publisher={IEEE}
}

@inproceedings{Sharafaldin2018TowardGA,
  title={Toward Generating a New Intrusion Detection Dataset and Intrusion Traffic Characterization},
  author={Iman Sharafaldin and Arash Habibi Lashkari and Ali A. Ghorbani},
  booktitle={International Conference on Information Systems Security and Privacy},
  year={2018},
  url={https://api.semanticscholar.org/CorpusID:4707749}
}

@misc{cicids2018,
  title        = {CSE-CIC-IDS2018 on AWS},
  author       = {{Canadian Institute for Cybersecurity}},
  year         = {2018},
  howpublished = {\url{https://www.unb.ca/cic/datasets/ids-2018.html}},
}

@inproceedings{schroff2015facenet,
  title={FaceNet: A unified embedding for face recognition and clustering},
  author={Schroff, Florian and others},
  booktitle={Proceedings of the IEEE conference on computer vision and pattern recognition},
  pages={815--823},
  year={2015}
}

@inproceedings{moustafa2015unsw,
  title={UNSW-NB15: a comprehensive data set for network intrusion detection systems (UNSW-NB15 network data set)},
  author={Moustafa, Nour and Slay, Jill},
  booktitle={Military Communications and Information Systems Conference (MilCIS)},
  pages={1--6},
  year={2015},
  organization={IEEE}
}

@article{wang2024comprehensive,
  title={A comprehensive survey of continual learning: theory, method and application},
  author={Wang, Liyuan and Zhang, Xingxing and Su, Hang and Zhu, Jun},
  journal={IEEE Transactions on Pattern Analysis and Machine Intelligence},
  year={2024},
  publisher={IEEE}
}

@misc{kim2023openworld,
  title={Open-World Continual Learning: Unifying Novelty Detection and Continual Learning},
  author={Gyuhak Kim and Changnan Xiao and Tatsuya Konishi and Zixuan Ke and Bing Liu},
  year={2023},
  eprint={2304.10038},
  archivePrefix={arXiv},
  primaryClass={cs.LG},
  url={https://arxiv.org/abs/2304.10038}
}

@book{han2011data,
  title={Data Mining: Concepts and Techniques},
  author={Han, Jiawei and Kamber, Micheline and Pei, Jian},
  year={2011},
  publisher={Elsevier}
}

@inproceedings{praucDavid06,
  author = {Davis, Jesse and Goadrich, Mark},
  title = {The relationship between Precision-Recall and ROC curves},
  year = {2006},
  isbn = {1595933832},
  publisher = {Association for Computing Machinery},
  address = {New York, NY, USA},
  url = {https://doi.org/10.1145/1143844.1143874},
  doi = {10.1145/1143844.1143874},
  booktitle = {Proceedings of the 23rd International Conference on Machine Learning},
  pages = {233–240},
  numpages = {8},
  location = {Pittsburgh, Pennsylvania, USA},
  series = {ICML '06}
}

@inproceedings{gungor2024roldef,
  title={ROLDEF: RObust Layered DEFense for Intrusion Detection Against Adversarial Attacks},
  author={Gungor, Onat and Rosing, Tajana and Aksanli, Baris},
  booktitle={2024 Design, Automation \& Test in Europe Conference \& Exhibition (DATE)},
  pages={1--6},
  year={2024},
  organization={IEEE}
}

@inproceedings{aljundi2022continual,
  title={Continual novelty detection},
  author={Aljundi, Rahaf and Reino, Daniel Olmeda and Chumerin, Nikolay and Turner, Richard E},
  booktitle={Conference on Lifelong Learning Agents},
  pages={1004--1025},
  year={2022},
  organization={PMLR}
}

@article{zhang2024ssf,
  title        = {Continual Learning with Strategic Selection and Forgetting for Network Intrusion Detection},
  author       = {Zhang, Xinchen and Zhao, Running and Jiang, Zhihan and Chen, Handi and Ding, Yulong and Ngai, Edith C. H. and Yang, Shuang-Hua},
  journal      = {arXiv preprint arXiv:2412.16264},
  year         = {2024},
  url          = {https://arxiv.org/abs/2412.16264},
  doi          = {10.48550/arXiv.2412.16264}
}

@article{zhang2024aocids,
  title        = {AOC-IDS: Autonomous Online Framework with Contrastive Learning for Intrusion Detection},
  author       = {Zhang, Xinchen and Zhao, Running and Jiang, Zhihan and Sun, Zhicong and Ding, Yulong and Ngai, Edith C. H. and Yang, Shuang-Hua},
  journal      = {arXiv preprint arXiv:2402.01807},
  year         = {2024},
  url          = {https://arxiv.org/abs/2402.01807},
  doi          = {10.48550/arXiv.2402.01807}
}

@article{li2025safe,
  title={SAFE: Self-Supervised Anomaly Detection Framework for Intrusion Detection},
  author={Li, Elvin and Shang, Zhengli and Gungor, Onat and Rosing, Tajana},
  journal={arXiv preprint arXiv:2502.07119},
  year={2025}
}

@article{shlens2014tutorial,
  title={A tutorial on principal component analysis},
  author={Shlens, J.},
  journal={arXiv preprint arXiv:1404.1100},
  year={2014}
}

@article{li2025citadel,
  title        = {CITADEL: Continual Anomaly Detection for Enhanced Learning in IoT Intrusion Detection},
  author       = {Li, Elvin and Gungor, Onat and Shang, Zhengli and Rosing, Tajana},
  journal      = {arXiv preprint arXiv:2508.19450},
  year         = {2025},
  url          = {https://arxiv.org/abs/2508.19450},
  doi          = {10.48550/arXiv.2508.19450}
}

@article{Torabi2023AutoencoderIDS,
  title     = {Autoencoder-based anomaly detection for network intrusion detection systems},
  author    = {Torabi, Hamed and Ghahramani, Mohammad and Shafiee, Mohammad and Azimifar, Zohreh},
  journal   = {Cybersecurity},
  volume    = {6},
  number    = {1},
  pages     = {1--21},
  year      = {2023},
  publisher = {SpringerOpen},
  doi       = {10.1186/s42400-022-00134-9},
  url       = {https://cybersecurity.springeropen.com/articles/10.1186/s42400-022-00134-9}
}

@article{Faber_2024,
  title={Lifelong Continual Learning for Anomaly Detection: New Challenges, Perspectives, and Insights},
  volume={12},
  ISSN={2169-3536},
  url={http://dx.doi.org/10.1109/ACCESS.2024.3377690},
  DOI={10.1109/access.2024.3377690},
  journal={IEEE Access},
  publisher={Institute of Electrical and Electronics Engineers (IEEE)},
  author={Faber, Kamil and Corizzo, Roberto and Sniezynski, Bartlomiej and Japkowicz, Nathalie},
  year={2024},
  pages={41364–41380}
}

@misc{yu2023scale,
  title={SCALE: Online Self-Supervised Lifelong Learning without Prior Knowledge},
  author={Xiaofan Yu and Yunhui Guo and Sicun Gao and Tajana Rosing},
  year={2023},
  eprint={2208.11266},
  archivePrefix={arXiv},
  primaryClass={cs.LG}
}

@article{macas2022survey,
  title={A survey on deep learning for cybersecurity: Progress, challenges, and opportunities},
  author={Macas, Mayra and Wu, Chunming and Fuertes, Walter},
  journal={Computer Networks},
  volume={212},
  pages={109032},
  year={2022},
  publisher={Elsevier}
}

@article{kozik2019balanced,
  title={Balanced Efficient Lifelong Learning (B-ELLA) for Cyber Attack Detection.},
  author={Kozik, Rafal and others},
  journal={J. Univers. Comput. Sci.},
  volume={25},
  number={1},
  pages={2--15},
  year={2019}
}

@inproceedings{channappayya2024augmented,
 author = {amalapuram, suresh kumar and Channappayya, Sumohana and Tamma, Bheemarjuna Reddy},
 booktitle = {Advances in Neural Information Processing Systems},
 editor = {A. Oh and T. Naumann and A. Globerson and K. Saenko and M. Hardt and S. Levine},
 pages = {17156--17169},
 publisher = {Curran Associates, Inc.},
 title = {Augmented Memory Replay-based Continual Learning Approaches for Network Intrusion Detection},
 url = {https://proceedings.neurips.cc/paper_files/paper/2023/file/3755a02b1035fbadd5f93a022170e46f-Paper-Conference.pdf},
 volume = {36},
 year = {2023}
}

@INPROCEEDINGS{fuhrman2025,
  author={Fuhrman, Sean and Gungor, Onat and Rosing, Tajana},
  booktitle={2025 62nd ACM/IEEE Design Automation Conference (DAC)}, 
  title={CND-IDS: Continual Novelty Detection for Intrusion Detection Systems}, 
  year={2025},
  volume={},
  number={},
  pages={1-7},
  doi={10.1109/DAC63849.2025.11132541}}

@article{FABER_VLAD,
title = {VLAD: Task-agnostic VAE-based lifelong anomaly detection},
journal = {Neural Networks},
volume = {165},
pages = {248-273},
year = {2023},
issn = {0893-6080},
doi = {https://doi.org/10.1016/j.neunet.2023.05.032},
url = {https://www.sciencedirect.com/science/article/pii/S0893608023002733},
author = {Kamil Faber and Roberto Corizzo and Bartlomiej Sniezynski and Nathalie Japkowicz},
}

@online{aag_cybercrime_stats,
  title     = {The Latest Cyber Crime Statistics},
  author    = {{AAG IT Services}},
  year      = {2025},
  url       = {https://aag-it.com/the-latest-cyber-crime-statistics/},
  note      = {Accessed: Dec. 4, 2025}
}

@inproceedings{
  madaan2022representational,
  title={Representational Continuity for Unsupervised Continual Learning},
  author={Divyam Madaan and Jaehong Yoon and Yuanchun Li and Yunxin Liu and Sung Ju Hwang},
  booktitle={International Conference on Learning Representations},
  year={2022},
  url={https://openreview.net/forum?id=9Hrka5PA7LW}
}

@article{de2021continual,
  title={A continual learning survey: Defying forgetting in classification tasks},
  author={De Lange, Matthias and others},
  journal={IEEE transactions on pattern analysis and machine intelligence},
  volume={44},
  number={7},
  pages={3366--3385},
  year={2021},
  publisher={IEEE}
}

@article{verwimp2023continual,
  title={Continual Learning: Applications and the Road Forward},
  author={Verwimp, Eli and others},
  journal={arXiv preprint arXiv:2311.11908},
  year={2023}
}

@article{verkerken2022towards,
  title={Towards model generalization for intrusion detection: Unsupervised machine learning techniques},
  author={Verkerken, Miel and D’hooge, Laurens and Wauters, Tim and Volckaert, Bruno and De Turck, Filip},
  journal={Journal of Network and Systems Management},
  volume={30},
  pages={1--25},
  year={2022},
  publisher={Springer}
}

@inproceedings{ruvolo2013ella,
  title={ELLA: An efficient lifelong learning algorithm},
  author={Ruvolo, Paul and Eaton, Eric},
  booktitle={International conference on machine learning},
  pages={507--515},
  year={2013},
  organization={PMLR}
}

@article{liu2019machine,
  title={Machine learning and deep learning methods for intrusion detection systems: A survey},
  author={Liu, Hongyu and Lang, Bo},
  journal={applied sciences},
  volume={9},
  number={20},
  pages={4396},
  year={2019},
  publisher={mdpi}
}

@article{liao2013intrusion,
  title={Intrusion detection system: A comprehensive review},
  author={Liao, Hung-Jen and others},
  journal={Journal of Network and Computer Applications},
  volume={36},
  number={1},
  pages={16--24},
  year={2013},
  publisher={Elsevier}
}

@inproceedings{amalapuram2022continual,
  title={Continual learning for anomaly based network intrusion detection},
  author={Amalapuram, Suresh Kumar and others},
  booktitle={2022 14th International Conference on COMmunication Systems \& NETworkS (COMSNETS)},
  pages={497--505},
  year={2022},
  organization={IEEE}
}

@inproceedings{oikonomou2023multi,
  title={A Multi-Class Intrusion Detection System Based on Continual Learning},
  author={Oikonomou, Chrysoula and others},
  booktitle={2023 IEEE International Conference on Cyber Security and Resilience (CSR)},
  pages={86--91},
  year={2023},
  organization={IEEE}
}

@article{prasath2022analysis,
  title={Analysis of continual learning models for intrusion detection system},
  author={Prasath, Sai and others},
  journal={IEEE Access},
  volume={10},
  pages={121444--121464},
  year={2022},
  publisher={IEEE}
}

@inproceedings{gungor2024rigorous,
  title={Rigorous Evaluation of Machine Learning-Based Intrusion Detection Against Adversarial Attacks},
  author={Gungor, Onat and Li, Elvin and Shang, Zhengli and Guo, Yutong and Chen, Jing and Davis, Johnathan and Rosing, Tajana},
  booktitle={2024 IEEE International Conference on Cyber Security and Resilience (CSR)},
  pages={152--158},
  year={2024},
  organization={IEEE}
}

@article{karn2021learning,
  title={Learning without forgetting: A new framework for network cyber security threat detection},
  author={Karn, Rupesh Raj and Kudva, Prabhakar and Elfadel, Ibrahim M},
  journal={IEEE Access},
  volume={9},
  pages={137042--137062},
  year={2021},
  publisher={IEEE}
}

@article{emerging2023kumar,
  title={Emerging Threats in Cybersecurity: A Review Article},
  volume={1},
  url={https://bluemarkpublishers.com/index.php/IJANS/article/view/2},
  number={1},
  journal={International Journal of Applied and Natural Sciences},
  author={Kumar, Iyana},
  year={2023},
  month={Jul.},
  pages={01–08}
}

@article{scikit-learn,
  title={Scikit-learn: Machine Learning in Python},
  author={Pedregosa, Fabian and Varoquaux, Ga{\"e}l and Gramfort, Alexandre and Michel, Vincent
          and Thirion, Bertrand and Grisel, Olivier and Blondel, Mathieu and Prettenhofer, Peter
          and Weiss, Ron and Dubourg, Vincent and Vanderplas, Jake and Passos, Alexandre 
          and Cournapeau, David and Brucher, Matthieu and Perrot, Matthieu and Duchesnay, {\'E}douard},
  journal={Journal of Machine Learning Research},
  volume={12},
  pages={2825--2830},
  year={2011}
}

@inproceedings{chaudhry2019tiny,
  title={Continual Learning with Tiny Episodic Memories},
  author={Chaudhry, Arslan and Ranzato, Marc'Aurelio and Rohrbach, Marcus and Elhoseiny, Mohamed},
  booktitle={Proceedings of the 36th International Conference on Machine Learning},
  pages={1286--1295},
  year={2019}
}

@article{gupta2023surveyIDS,
  author = {Gupta, Neha and Jindal, Vinita and Bedi, Punam},
  title = {A Survey on Intrusion Detection and Prevention Systems},
  year = {2023},
  publisher = {Springer-Verlag},
  address = {Berlin, Heidelberg},
  journal = {SN Comput. Sci.},
  month = jun,
  numpages = {28}
}

@article{al2021x,
  title={X-IIoTID: A connectivity-agnostic and device-agnostic intrusion data set for industrial Internet of Things},
  author={Al-Hawawreh, Muna and others},
  journal={IEEE Internet of Things Journal},
  volume={9},
  number={5},
  pages={3962--3977},
  year={2021},
  publisher={IEEE}
}

@inproceedings{zong2018deep,
  title={Deep autoencoding gaussian mixture model for unsupervised anomaly detection},
  author={Zong, Bo and Song, Qi and Min, Martin Renqiang and Cheng, Wei and Lumezanu, Cristian and Cho, Daeki and Chen, Haifeng},
  booktitle={International conference on learning representations},
  year={2018}
}

@ARTICLE{ashfahani2023unsupervised,
  author={Ashfahani, Andri and Pratama, Mahardhika},
  journal={IEEE Transactions on Neural Networks and Learning Systems},
  title={Unsupervised Continual Learning in Streaming Environments},
  year={2023},
  volume={34},
  number={12},
  pages={9992-10003},
  doi={10.1109/TNNLS.2022.3163362}
}

@inproceedings{rios2022incdfm,
  title={incdfm: Incremental deep feature modeling for continual novelty detection},
  author={Rios, Amanda and Ahuja, Nilesh and Ndiour, Ibrahima and Genc, Utku and Itti, Laurent and Tickoo, Omesh},
  booktitle={European Conference on Computer Vision},
  pages={588--604},
  year={2022},
  organization={Springer}
}

@ARTICLE{lwf2019Li,
  author={Li, Zhizhong and Hoiem, Derek},
  journal={IEEE Transactions on Pattern Analysis and Machine Intelligence},
  title={Learning without Forgetting},
  year={2018},
  volume={40},
  number={12},
  pages={2935-2947},
  doi={10.1109/TPAMI.2017.2773081}
}

@misc{zolanvari2021wustl,
  author = {Zolanvari, M. and others},
  title = {WUSTL-IIOT-2021 Dataset for IIoT Cybersecurity Research},
  howpublished = {Washington University in St. Louis, USA},
  year = {2021},
  month = {oct},
  note = {\url{http://www.cse.wustl.edu/~jain/iiot2/index.html}}
}

@article{diaz2018don,
  title={Don't forget, there is more than forgetting: new metrics for Continual Learning},
  author={D{\'\i}az-Rodr{\'\i}guez, Natalia and Lomonaco, Vincenzo and Filliat, David and Maltoni, Davide},
  journal={arXiv preprint arXiv:1810.13166},
  year={2018}
}

@inproceedings{sculley2010web,
  title={Web-scale k-means clustering},
  author={Sculley, David},
  booktitle={Proceedings of the 19th International Conference on World Wide Web},
  pages={1177--1178},
  year={2010},
  organization={ACM},
  doi={10.1145/1772690.1772862}
}

@article{garcia2009anomaly,
title = {Anomaly-based network intrusion detection: Techniques, systems and challenges},
journal = {Computers \& Security},
volume = {28},
number = {1},
pages = {18-28},
year = {2009},
issn = {0167-4048},
doi = {https://doi.org/10.1016/j.cose.2008.08.003},
url = {https://www.sciencedirect.com/science/article/pii/S0167404808000692},
author = {P. García-Teodoro and J. Díaz-Verdejo and G. Maciá-Fernández and E. Vázquez},
}






\end{document}